
\magnification=\magstep1
\vsize=8.5 true in
\hsize=6.5 true in
\tolerance 10000

\hyphenation{par-a-me-ters}
\baselineskip 18pt plus 1pt minus 1pt
\pageno=0
\vskip 0.3in
\centerline{\bf Constraints on cosmic strings due to}
\centerline{\bf black holes formed from collapsed cosmic string
loops}
\vskip 0.5in
\centerline{R. R. Caldwell$^\dagger$ and Evalyn
Gates$^{\dagger\ddagger}$}
\vskip 0.5in
\centerline{\dag NASA/Fermilab Astrophysics Center}
\centerline{Fermi National Accelerator Laboratory}
\centerline{P.O. Box 500}
\centerline{Batavia, Illinois 60510-0500}
\vskip 0.2in
\centerline{\ddag University of Chicago}
\centerline{5640 S. Ellis Avenue}
\centerline{Chicago, Illinois 60637}
\vskip 0.2in

\centerline{\it May, 1993}
\vfil
\baselineskip 12pt plus 1pt minus 1pt

\centerline{\bf ABSTRACT}
{\noindent\narrower
The cosmological features of primordial black holes
formed from collapsed cosmic string loops are studied.
Observational restrictions on a population of
primordial black holes are used to restrict
$f$, the fraction of cosmic string
loops which collapse to form black holes,
and $\mu$, the cosmic string mass-per-unit-length.
Using a realistic model of cosmic strings, we find the
strongest restriction on the parameters $f$ and $\mu$
is due to
the energy density in $100 MeV$ photons radiated by the
black holes.
We also find that inert black hole remnants cannot serve
as the dark matter.
If earlier, crude estimates of $f$ are reliable, our
results severely restrict $\mu$, and therefore
limit the viability of the cosmic string
large-scale structure scenario.

\vfil
\eject

\expandafter\ifx\csname phyzzx\endcsname\relax\else
 \errhelp{Hit <CR> and go ahead.}
 \errmessage{PHYZZX macros are already loaded or input. }
 \endinput \fi
\catcode`\@=11 
\def\spacecheck#1{\dimen@=\pagegoal\advance\dimen@ by -\pagetotal
   \ifdim\dimen@<#1 \ifdim\dimen@>0pt \vfil\break \fi\fi}
\newskip\chapterskip         \chapterskip=\bigskipamount
\newskip\headskip            \headskip=8pt plus 3pt minus 3pt
\newdimen\referenceminspace  \referenceminspace=25pc
\font\fourteenrm=cmr10 scaled\magstep2
\def\refitem#1{\par \hangafter=0 \hangindent=\refindent
\Textindent{#1}}
\def\Textindent#1{\noindent\llap{#1\enspace}\ignorespaces}
\def\subspaces@t#1:#2;{
      \baselineskip = \normalbaselineskip
      \multiply\baselineskip by #1 \divide\baselineskip by #2
      \lineskip = \normallineskip
      \multiply\lineskip by #1 \divide\lineskip by #2
      \lineskiplimit = \normallineskiplimit
      \multiply\lineskiplimit by #1 \divide\lineskiplimit by #2
      \parskip = \normalparskip
      \multiply\parskip by #1 \divide\parskip by #2
      \abovedisplayskip = \normaldisplayskip
      \multiply\abovedisplayskip by #1 \divide\abovedisplayskip by #2
      \belowdisplayskip = \abovedisplayskip
      \abovedisplayshortskip = \normaldispshortskip
      \multiply\abovedisplayshortskip by #1
        \divide\abovedisplayshortskip by #2
      \belowdisplayshortskip = \abovedisplayshortskip
      \advance\belowdisplayshortskip by \belowdisplayskip
      \divide\belowdisplayshortskip by 2
      \smallskipamount = \skipregister
      \multiply\smallskipamount by #1 \divide\smallskipamount by #2
      \medskipamount = \smallskipamount \multiply\medskipamount by 2
      \bigskipamount = \smallskipamount \multiply\bigskipamount by 4
}
%
%
%
%
\newcount\referencecount     \referencecount=0
\newcount\lastrefsbegincount \lastrefsbegincount=0
\newif\ifreferenceopen       \newwrite\referencewrite
\newif\ifrw@trailer
\newdimen\refindent     \refindent=30pt
\def\NPrefmark#1{\attach{\scriptscriptstyle [ #1 ] }}
\let\PRrefmark=\attach
\def\refmark#1{\relax\ifPhysRev\PRrefmark{#1}\else\NPrefmark{#1}\fi}
\def\refend@{\refmark{\number\referencecount}}
\def\refend{\refend@{}\space }
\def\refsend{\refmark{\count255=\referencecount
   \advance\count255 by-\lastrefsbegincount
   \ifcase\count255 \number\referencecount
   \or \number\lastrefsbegincount,\number\referencecount

   \else \number\lastrefsbegincount-\number\referencecount \fi}\space
}
\def\refitem#1{\par \hangafter=0 \hangindent=\refindent
\Textindent{#1}}
\def\Ref{\rw@trailertrue\REF}
\def\ref{\Ref\?}

\def\REF#1{\r@fstart{#1}%
   \rw@begin{\the\referencecount.}\rw@end}
\def\REFS#1{\r@fstart{#1}%
   \lastrefsbegincount=\referencecount
   \rw@begin{\the\referencecount.}\rw@end}
\def\r@fstart#1{\chardef\rw@write=\referencewrite
\let\rw@ending=\refend@
   \ifreferenceopen \else \global\referenceopentrue
   \immediate\openout\referencewrite=referenc.txa
   \toks0={\catcode`\^^M=10}\immediate\write\rw@write{\the\toks0} \fi
   \global\advance\referencecount by 1 \xdef#1{\the\referencecount}}
{\catcode`\^^M=\active %
 \gdef\rw@begin#1{\immediate\write\rw@write{\noexpand\refitem{#1}}%
   \begingroup \catcode`\^^M=\active \let^^M=\relax}%
 \gdef\rw@end#1{\rw@@end #1^^M\rw@terminate \endgroup%
   \ifrw@trailer\rw@ending\global\rw@trailerfalse\fi }%
 \gdef\rw@@end#1^^M{\toks0={#1}\immediate\write\rw@write{\the\toks0}%
   \futurelet\n@xt\rw@test}%
 \gdef\rw@test{\ifx\n@xt\rw@terminate \let\n@xt=\relax%
       \else \let\n@xt=\rw@@end \fi \n@xt}%
}
\let\rw@ending=\relax
\let\rw@terminate=\relax

\def\Closeout#1{\toks0={\catcode`\^^M=5}\immediate\write#1{\the\toks0
}%
   \immediate\closeout#1}

\REF\HAWKINGA{
S. W. Hawking, Phys. Lett. {\bf B231}, 237 (1989).}
\REF\HAWKINGB{
S. W. Hawking, Phys. Lett. {\bf B246}, 36 (1990).}
\REF\BARRABES{
C. Barrabes, Class. Quantum Grav. {\bf 8} L199 (1991).}
\REF\POLNAREV{
Alexander Polnarev and Robert Zembowicz, Phys. Rev. {\bf D43}, 1106
(1991).}
\REF\VILENKIN{
Alexander Vilenkin and Jaume Garriga, {\it to appear} PRD (1993).}
\REF\CARRA{
B. J. Carr, Ap. J. {\bf 201}, 1 (1975).}
\REF\PAGEB{
Don N. Page and S. W. Hawking, Ap. J. {\bf 206}, 1 (1976).}
\REF\CARRB{
Bernard J. Carr, Ap. J. {\bf 206}, 8 (1976).}
\REF\CARRC{
B. J. Carr, ``Black hole evaporations and their cosmological
consequences'' in
{\it Quantum Gravity}, Markov and West, eds. (1980). }
\REF\MACGIBBONB{
Jane H. MacGibbon and B. J. Carr, Ap. J. {\bf 371}, 447 (1991).}
\REF\BARROW{
John D. Barrow, Edmund J. Copeland, and Andrew R. Liddle,
Phys. Rev. {\bf D46}, 645 (1992).}
\REF\LIDSEY{
B. J. Carr and James E. Lidsey, ``Primordial Black Holes and
Generalized
Constraints
on Chaotic Inflation'',  Phys. Rev {\bf D}, {\it to appear} (1993).}
\REF\QUASHA{
R. J. Scherrer, J. M. Quashnock, D. N. Spergel, and W. H. Press,
Phys. Rev. {\bf D42}, 1908 (1990).}
\REF\BB{
David  Bennett and Francois Bouchet, Phys. Rev {\bf D41}, 2408
(1990).}
\REF\AS{
B. Allen and E. P. S. Shellard, Phys. Rev. Lett. {\bf 64}, 119
(1990).}
\REF\AT{
A. Albrecht and N. Turok, Phys. Rev. {\bf D40}, 973 (1989).}
\REF\QUASHB{
J. M. Quashnock and D. N. Spergel, Phys. Rev. {\bf D42}, 2505
(1990).}
\REF\GARRIGA{
Jaume Garriga and Maria Sakellariadou, TUTP-93-4 (1993).}
\REF\CG{
R. R. Caldwell and E. Gates, {\it work in progress} (1993).}
\REF\KINKYMODEL{
B. Allen and R. R. Caldwell, Phys. Rev. Lett {\bf 65}, 1705 (1990);
Mark Hindmarsh, Phys. Lett. {\bf B251}, 28 (1990);
T. W. B. Kibble and E. Copeland, in {\it The Birth and Early
Evolution of Our
Universe},
Proceedings of the Nobel Symposium 79, edited by
B. S. Skagerstam (World Scientific) (1990);
B. Allen and R. R. Caldwell, Phys. Rev. {\bf D43}, R2457 (1991);
B. Allen and R. R. Caldwell, Phys. Rev. {\bf D43}, 3173 (1991);
J. Quashnock and Tsvi Piran, Phys. Rev. {\bf D43}, R3785 (1991).}
\REF\CONSTRAINTS{
R. R. Caldwell and B. Allen, Phys. Rev. {\bf D45}, 3447 (1992).}
\REF\PAGEA{
Don N. Page, Phys. Rev. {\bf D13}, 198 (1976).}
\REF\REMNANT{
M. A. Markov in {\it Proc. 2nd Seminar in
Quantum Gravity}, eds. M. A. Markov and P. C. West, 1 (Plenum, New
York,
1984).}
\REF\MACGIBBONA{
J. H. MacGibbon, Nature {\bf 329}, 308 (1987).}
\REF\FLUIDFRICTION{
Alexander Vilenkin, Phys. Rev. {\bf D43}, 1060 (1991).}
\REF\FICHTEL{
C. E. Fichtel,R. C. Hartman, D. A. Kniffen,
D. J. Thompson, G. F. Bignami, H. \"Ogelman, M. E. \"Ozel, and
T. T\"umer, Ap. J. {\bf 198} 163 (1975).}
\REF\MACGIBBONC{
F. Halzen, E. Zas, J. H. MacGibbon, and T. C. Weekes,
Nature {\bf 353}, 807 (1991).}

\centerline{\bf I. Introduction}

The cosmic string scenario for the formation
of large scale structure has many observable features.
Primarily, cosmic strings may serve to produce perturbations
to the cosmological fluid of the necessary magnitude and distribution
to seed the formation of galaxies and clusters, as observed today.
Cosmic strings leave an observational signature
through these perturbations, as well as through the
emission of gravitational radiation.
Broadly, then, there are two areas of cosmic string research.
These are studies of the large-scale structure produced
by cosmic strings, and tests of the compatibility of cosmic
strings with cosmological observations.
Such tests focus, for example, on the anisotropies produced
by cosmic strings in the microwave background and the noise in pulsar
timing
due to the cosmic string stochastic gravitational wave background.
Ultimately, the test of compatibility results in a restriction
on $\mu$, the mass-per-unit-length and sole free parameter
in the cosmic string model. In this report, we will examine
the restrictions on black holes formed from collapsed cosmic string
loops.

It is well known that a sufficiently smooth, circular cosmic string
loop may collapse to form a black hole
[\HAWKINGA,\HAWKINGB,\BARRABES,\POLNAREV,\VILENKIN].
During the evolution of a network of cosmic strings, some cosmic
string
loops may collapse to form black holes.
In this case, the observational restrictions on a population of
primordial
black holes
may be used to restrict such a cosmic string scenario.

The study of primordial black holes has been vigorously carried out
in, for example,
[\CARRA,\PAGEB,\CARRB,\CARRC,\MACGIBBONB,\BARROW,\LIDSEY].
We will take advantage of this work in applying
constraints to a population of black holes formed from collapsed
cosmic
string loops. In turn, we will place restrictions on the cosmic
string
network. {\it In this paper, we will find observational
restrictions on the cosmic string scenario from cosmic string
loops which collapse to form black holes.}

The organization of this paper is as follows. In section
II we will summarize previous efforts to estimate the fraction
$f$ of cosmic string loops which collapse to form black holes.
In section III we will present the models of cosmic strings and
black hole evaporation used to calculate the energy density in
black holes and black hole radiation.
In section IV we will present the observational constraints on
a population of black holes formed from collapsed cosmic string
loops. We will conclude in section V with a
restriction on the parameters $f$ and $\mu$.

\vskip 0.2in
\centerline{\bf II. Collapse of Cosmic String Loops}
\centerline{\bf to form Black Holes}

A cosmic string which contracts under its own tension
to a size smaller than its Schwarzschild radius will form a black
hole.
In this section, we will conduct a brief review of
the analysis of this phenomena. We will present a naive
estimate of the probability that a realistic cosmic string loop
will collapse to form a black hole. While no conclusive
work has been carried out to determine this fraction,
our naive estimate will serve as a rough guide for
the cosmological analysis in the succeding sections.

The phenomena by which a cosmic string loop collapses to form
a black hole may be best understood by examining a simple case.
We will consider a perfectly circular, planar cosmic string loop
of mass $m$. The string equations of motion dictate that
such a loop will expand and contract under its own tension,
with a maximum radius $R_{max} = m/2 \pi \mu$.
When it contracts under its own tension
to within its Schwarzschild radius $R_S = 2 G m = 4 \pi G \mu R$,
it will form a black hole.
(We use units $\hbar = c = 1$ and $G = m_{planck}^{-2}$.)
A loop may never contract to within its Schwarzschild
radius, however, if it is sufficiently non-circular.
As well, a loop with a Schwarzschild radius comparable to its
thickness may dissipate, by radiating the quanta trapped in the
string,
before a black hole may form. Thus, not all cosmic string loops
collapse to form black holes; in fact we expect only a small fraction
to do so.

We are interested in determining which loops formed by a
realistic cosmic string network collapse to form black holes.
Simplifying this problem, we ask what fraction $f$ of cosmic string
loops
collapse to form black holes. We will ultimately find that
observational
restrictions on black holes formed from collapsed cosmic string loops
will
depend linearly on this fraction $f$.  In this study we will not be
able to
conclusively determine $f$. After reviewing past work on the
properties
and behavior of realistic cosmic string loops, however, we will be
able to
determine the
relevant properties that effect this fraction.  In addition, previous
attempts
to determine this fraction $f$, along with reasonable assumptions
about the
loop population, will be shown to indicate a rough value for $f$.  If
these
estimates are
reasonable, we may be able to place severe restrictions on cosmic
string
scenarios.

The initial investigation of black holes formed from cosmic string
loops
was carried out by Hawking [\HAWKINGA]. Considering a scenario where
the loop
would
have a similar probability of collapsing to form a black hole in each
oscillation
period, he proposed that
the fraction $f$ is a function of the mass per unit length of the
loop and
the number of kinks ($n$) on a string loop.  His expression for $f$,
however,
depends exponentially on
$n$. Numerical simulations of cosmic string evolution suggest that a
reasonable
range for $n$ is given by $n \sim 2-5$ [\QUASHA],
which corresponds to, for $\mu = 10^{-6}$,
$f \sim 1 - 10^{-36}$.  Thus, unless a more accurate
determination of $n$ is achieved, this approach does not provide a
conclusive
estimate for $f$ that may be useful for constraining cosmic string
scenarios.

A numerical analysis of several families of parameterized
loop configurations was carried out by Polnarev and
Zembowicz [\POLNAREV]. They examined the fraction of parameter space
for which
Burden and Kibble-Turok families of loops collapse to form black
holes.
These families model loops which contain cusps, and may not
necessarily
be representative of realistic loops.
Depending on the measure assigned to the configuration parameter
space, they
found the fraction may lie in the range $f \sim 10^{-9}-10^{-15}$.
For loops with few kinks, this range of values seems to be roughly in
accord with Hawking's estimates.

There are two major features of a cosmic string loop which may
determine whether the loop will collapse to form a black hole.
These are, roughly, the large (underlying loop configuration or
shape)
and small (kinks and cusps) scale features of the loop. One may
attempt to determine the fraction of loops
collapsing to black holes by asking the following questions:
(i) what fraction of realistic loops possess an underlying
configuration which would lead to the formation of a black hole,
and (ii) what fraction of these loops possess kinks and cusps
(fluctuations) small enough that a black hole may still form.
Previous numerical simulations of cosmic string evolution
[\QUASHA,\BB,\AS,\AT] provide some insight into the relevant issues.

The study of the effects of the gravitational back-reaction
on the evolution of cosmic string loops indicates that
the gravitational back-reaction will set the minimum
scale of structures on long strings. These structures are the
predecessors of parent loops which are chopped off the long strings.
The parent loops then undergo fragmentation and rapidly evolve
towards
simple, non-intersecting
configurations, containing on the order of $2 - 5$ kinks. Quashnock
and
Spergel [\QUASHB] found that the
kinks and small scale structure on string loops rapidly decay, and
the
loops then oscillate in a self-similar manner. Cusps, however, are
not
suppressed by the gravitational back-reaction and persist throughout
the evolution of the loop. Thus, except for configurations that
contain
cusps, the underlying shape of a realistic loop is dominated by low
mode
or long wavelength oscillations.

The work of Garriga and Vilenkin [\GARRIGA] considered nucleated
cosmic string loops, which may collapse to form black holes.
The behavior of classical fluctuations on the loops were
examined, and it was shown that
while transverse perturbations maintain constant amplitude
as the cosmic string loop contracts,
radial perturbations shrink by a factor of the perturbation mode
number.
Since the loops are dominated by low mode oscillations, the maximum
allowed
amplitude of a perturbation that is consistent with collapse to a
black hole
is of order $R_s$.  That is, the maximum tolerable fluctuation which
will not
prevent the formation of a black hole contains only $\sim G \mu$ of
the total
loop energy.
However, while we expect most of the daughter (non-self-intersecting)
loops
to be relatively simple, we currently have no information regarding
the
frequency or distribution of fluctuations
and loop configurations.

We may nevertheless attempt to estimate the fraction $f$
using Hawking's approach, incorporating the work of
[\QUASHA] on the properties of realistic cosmic string loops.
We consider the properties of
stable, non-self-intersecting loops,
and assume that the number of kinks on a loop
is the dominant factor in determining whether such a
loop will collapse to form a black hole. Note that since these
loops oscillate in a self-similar manner, we are concerned only
with whether a given configuration will immediately collapse to form
a
black hole. We do not integrate this probability over the number
of oscillations in the black hole lifetime, as Hawking did in his
original
calculation.
Thus, we then integrate Hawking's expression for the number
of loops which collapse {\it immediately} to form black holes
over the distribution of the number of kinks on stable
daughter loops.  This integral is dominated by the contribution
from loops with two kinks. We find $$f \sim 10^{-12}.$$
This estimate is roughly in the same range as that given in the
work by Zembowicz and Polnarev.
We must stress that although this is just a rough estimate,
we may use this fraction as a guide
for our study of the observational constraints in
the following section.

A careful determination of $f$ will be necessary to conclusively
evaluate the observational constraints on cosmic strings.
Such a study will be the focus of a future work [\CG].
In the meantime, we will adopt the working hypothesis
that {\it a fraction $f$ of realistic loops are smooth enough at the
time they are chopped off the cosmic string network that they
may immediately collapse to form black holes.}
We may now proceed to evaluate the observational
constraints on black holes formed from collapsed cosmic string loops.

\vskip 0.2in
\centerline{\bf III. Production and Evolution of Black Holes}
\centerline{\bf from the Collapse of Cosmic String Loops}

The properties of a population of black holes formed from collapsed
cosmic string loops are well specified by the properties of the
cosmic string network and by the properties of quantum mechanical
evaporation by a black hole. In this section we will first
present the model of cosmic strings used in this study, focusing on
those aspects relevant to the population of black holes produced
from collapsed cosmic string loops. Second, we will
describe how the quantum mechanical decay of black holes
is incorporated into the cosmic string scenario.
Third, we will outline
the calculation of the physical properties of the population of
black holes necessary to make contact with cosmological observations,
and subsequently restrict the cosmic string model.

We use the ``one-scale'' model of kinky cosmic strings. The
properties
of this model have been well described by [\KINKYMODEL].
We will repeat the necessary elements of this model.
\item{i.}
The background cosmology is a
spatially flat FRW spacetime with scale factor $a(t) \propto t^{1/2}$
in the radiation dominated era, and
$a(t) \propto t^{2/3}$ in the matter dominated era.
\item{ii.}
For simplicity, we will calculate physical quantities
in a fiducial, physical volume $V(t) = a^3(t) r^3$
where $r$ is an arbitrary coordinate length.
\item{iii.}
We define loops to be closed cosmic strings
formed, with an initial size $L(t) = \alpha l(t)$, through the
intercommutation of long strings, where $l(t)$ is the horizon radius.
All other cosmic string is contained in long strings.
\item{iv.}
Loops are considered to be non-self-intersecting.
\noindent{One may argue that a newly formed loop may self-intersect
and fragment at a rate proportional to the loop oscillation
frequency, producing smaller loops. This rate at which a loop
self-intersects, however, is much faster than both the rate at which
a
loop radiates gravitational waves and the expansion rate. Thus, we
are
justified in assuming that a loop fragments rapidly; we may consider
that $L(t)$ represents the size of the final, non-self-intersecting
loops.}
\item{v.}
The rate of loop formation is [\CONSTRAINTS]
$${d N_{loop} \over d t} = 4 {A \over \alpha} t^{-4} V(t)
\eqno(III.1)$$
where $A \approx 10$ gives the number of long, horizon-length cosmic
strings
present in a horizon volume, as determined by numerical simulations
[\BB,\AS].
The value $\alpha \approx 10^{-4}$ is given by the observational
bounds
on cosmic string gravitational radiation [\CONSTRAINTS].

We may now obtain the rate of black hole formation from collapsed
cosmic
string loops. Applying our hypothesis regarding the formation
of black holes from cosmic string loops to equation III.1, we find
$${d N_{bh} \over d t} = f {d N_{loop} \over d t}. \eqno(III.2)$$
This equation states that $N_{bh}$ black holes of
mass $m(t) = \alpha \mu l(t)$ were formed during the
time interval $t$ to $t + dt$. This equation is valid
for times $t \gg t_i$, where $t_i \sim \alpha^{-1} \mu^{-3/2}
t_{planck}$
gives the time at which the Schwarzschild radius of a newly formed
loop
is comparable to the thickness of the cosmic string.
Thus, the properties of the cosmic string network
determine the initial properties of the population of
black holes.

The cosmological evolution of a black hole is dominated
by quantum mechanical emission of a spectrum
of particles [\HAWKINGB]. This radiation, which has been well
investigated, is the most important
cosmological aspect of a black hole.
Thus, in order to follow the evolution of
the black holes formed from collapsed cosmic string loops
we need the black hole decay rate.
For a black hole of mass $m$, the decay into massless
particles is given by [\PAGEA]
$$ {d^2 m \over d \omega dt} =  \sum_i \Gamma(m,\omega,s) {\omega
\over
e^{8 \pi m \omega} \pm 1} \eqno(III.3)$$
where the sum is over all particle species $i$.
The $\pm$ refers to boson or fermion statistical weights,
and $\Gamma(m,\omega,s)$ is a dimensionless
function of the black hole mass, the radiated particle spin $s$,
and frequency $\omega$. We will be interested primarily
in the emission of photons, for which
$\Gamma(m,\omega,s) = 64 m^4 \omega^4 /9$.
Examining III.3, we see that a
black hole emits a burst of thermal radiation,
characterized by the black hole temperature, which is inversely
proportional to the black hole mass.
This emission will continue until the black hole has
completely evaporated away, or, as has been suggested
[\BARROW,\REMNANT,\MACGIBBONA], an inert Planck-mass object remains.
In such a case, the black hole evaporation rate will
truncate when $m \sim m_{planck}$. Adding the
black hole decay rate into our model, therefore, we have completely
specified
the cosmological evolution of a population of black holes.

We may now use the expressions for the rate of black hole formation
and the rate of black hole evaporation to calculate the energy
density
produced in black holes and black hole radiation.
The fraction of critical energy density in black holes is
$$ \Omega_{bh}(t) = {1 \over \rho_{crit}(t)} {1 \over V(t)}
\int_{t_i}^t dt' {d N_{bh} \over dt'} m(t',t). \eqno(III.4)$$
The limits of integration are from the time $t_i$ when the cosmic
string loops may first collapse to form black holes, to the present
time $t$.
Here, the function $m(t',t)$ gives the mass of a black hole formed at
time
$t'$ at a later time $t$. This function may be obtained by
integrating
the black hole evaporation rate III.3 over frequency and time,
applying
suitable boundary conditions. Examining III.4, the energy density in
black
holes at time $t$ is dominated by those surviving black holes which
formed earliest, as the rate of black hole production is a rapidly
decreasing function of time.
The fraction of critical energy density in black hole radiation, in
a logarithmic frequency interval, is
$${d \Omega_{bh rad}(t) \over d \ln{\omega}} = {1 \over
\rho_{crit}(t)}
{1 \over V(t)} \int_{t_i}^{t} dt' {d N_{bh} \over d t'}
\int_{t'}^{\tau(t') + t'} dt''  {\omega(t'') d^2 m \over d\omega(t'')
dt''}.
\eqno(III.5)$$
Here, $\tau(t')$ is the lifetime of a black hole,
and $\omega(t'') = \omega a(t)/a(t'')$
gives the relationship between the frequency as emitted at time
$t''$,
$\omega(t'')$, and the frequency observed at time $t$.
This power spectrum is dominated by the contribution
from black holes evaporating at the present time.
These expressions, III.4
and III.5 have been integrated numerically; the
results will be presented in section IV.

We will be interested, as well, in
constructing the function $\beta(m)$ in
order to evaluate the observational constraints
on this population of
black holes formed from collapsed cosmic string loops.
This function represents the fraction of critical energy density
in black holes formed during the time interval $t$ to $t+dt$.
$$\eqalign{\beta(m) =& {1 \over \rho_{crit}(t)}
{m(t) \over V(t)} d N_{bh} \cr
=& {256 \pi \over 3} A \mu f \cr} \eqno(III.6)$$
(Here, we have used $\rho_{crit}(t) = 3 t^{-2}/32 \pi$.)
It is not surprising that this function is really a constant.
The gross features of the cosmic string network scale
with the horizon radius: the gross features of the population
of black holes formed from collapsed cosmic string loops
scale with the horizon radius.
It is important to note that equation III.6
describes a population of black holes different
from the black holes described by the function $\beta(m)_{horizon}$
found in the primordial black hole literature (for example, see
[\MACGIBBONB]).
There, $\beta(m)_{horizon}$ represents the fraction
of critical energy density in black holes which
enter the horizon in the time interval $t$ to $t + dt$.
Such a black hole will be much smaller than the horizon
radius, as are the black holes formed from collapsed
cosmic string loops, at the later time $(2 \alpha \mu)^{-1} t$.
Therefore, to relate equation III.6 to the function
$\beta(m)_{horizon}$
found in the literature, we write
$$\eqalign{\beta(m) = & \beta(m)_{horizon}
{a(t/2 \alpha \mu) \over a(t)} \cr
\approx & (2 \alpha \mu)^{-1/2}
\beta(m)_{horizon}. \cr} \eqno(III.7)$$
Here, we have simply accounted for the growth in the black hole
energy density over the background radiation energy density
from the time the black hole enters the horizon to the
time that a black hole of the same mass would be formed from
a collapsed cosmic string loop.
The function $\beta$ will be used in section IV,
as has been used in [\MACGIBBONB,\LIDSEY],
to evaluate the restrictions on black holes.

We have apparently neglected to
consider the loops formed along with the long strings at the time
of the cosmological phase transition. These loops, as has been
recently shown in [\GARRIGA], may be smoothed by the friction with
the cosmological fluid [\FLUIDFRICTION]. The loops most smoothed by
the
friction,
however, have Schwarzschild radii smaller than the string thickness;
these loops will not collapse to form black holes. The remaining,
unsmoothed loops, which are larger than the horizon radius at time
$t_i$,
behave simply as long strings. Thus, we argue that we have considered
all loops which may collapse to form black holes, and may now proceed
to evaluate the observational constraints on black holes.

\vskip 0.2in
\centerline{\bf IV. Observational Constraints on Black Holes }
\centerline{\bf from Collapsed Cosmic String Loops}

The numerous observational restrictions on a population of primordial
black
holes are a direct consequence of the richness of the physics
of black hole evaporation. Through quantum mechanical decay,
a black hole will radiate all particle species.
Primordial black holes may be observed then through the
emitted particle spectra.
Consequently, the observation of spectra produced by
such black holes may serve to indicate exotic events which may
have taken place in the early universe.
Figuratively, the cosmic string
energy invested in black holes in the early universe
provides a return with observational consequences today.
In the following paragraphs we will present the
observational constraints on black holes formed by
collapsed cosmic string loops.
We will begin by evaluating equations III.4-5 for constraints on
the energy density in black hole photon radiation and remnants.
Next, we will use equations III.6-7 to evaluate more
observational constraints.
These constraints will be expressed in terms of a restriction on
$f$, using the preferred value $\mu = 10^{-6}$.
We will conclude this section with an
interpretation of the observational constraints on the
cosmic string parameters.

The strongest constraint on this population of black holes formed
from
collapsed cosmic string loops is due to the
$\gamma$ ray flux observed at $100 MeV$
[\FICHTEL,\MACGIBBONB,\MACGIBBONC].
We require that the fraction of critical energy
density in photons emitted by black holes, with
energy in a logarithmic interval at $100 MeV$,
be less than $\Omega_{\gamma} = 10^{-8} h^{-2}$.
Integrating equation III.5, we find
$$\eqalign{ {d \Omega_{bh \gamma}(t_0) \over d \ln{\omega}}
\mid_{\omega = 100 MeV}
= 10^{9} h^{-2} f \le & 10^{-8} h^{-2} \cr
\rightarrow f \le & 10^{-17} .\cr }\eqno(IV.1)$$
Black holes of mass $m \sim 10^{16} g$ (with a
lifetime $\sim 10^{17} s$) which evaporate today
serve as the dominant source of photons at this energy.

In the work of both Hawking [\HAWKINGA] and Polnarev and Zembowicz
[\POLNAREV]
the cosmological constraints on cosmic strings due to the  $\gamma$
rays
emitted by black holes formed from collapsed cosmic string loops were
evaluated. These
calculations used a very rough cosmic string model. Our work improves
upon
their results by implementing a realistic cosmic string model, as we
take
advantage of results from numerical simulations to determine the
average number
of long strings in a horizon volume and the size of newly formed
loops.
The improved limits are due to our improved model.

Further observational constraints on primordial black
holes, which are typically
stated as a restriction on the energy density in black holes
in a particular mass range, may be easily evaluated
analytically using $\beta$. We have taken advantage of the
literature [\MACGIBBONB,\LIDSEY], in which the restrictions on black
holes are
stated in terms of $\beta(m)_{horizon}$, which we may
simply convert into $\beta$ according to equation III.7.
In the following table, then, we list the observational
constraint, the restriction on $\beta$, and
the resultant limit on $f$, the fraction of collapsing cosmic string
loops.

\vskip 0.2in
\input tables

\begintable
\multispan{3}\tstrut\hfil Observational Restrictions on $f$
\hfil\crthick
diffuse $\gamma$ ray background 		| $\beta(m_{15}) \le
10^{-21}$ 	| $f \le
10^{-17}$ \cr
interstellar $e^{+}$ background		| $\beta(m_{15}) \le
10^{-21}$	| $f \le
10^{-17}$ \cr
interstellar $\bar p$ background	| $\beta(m_{15}) \le 10^{-21}$
	| $f \le
10^{-17}$ \cr
interstellar $e^{-}$ background		| $\beta(m_{15}) \le
10^{-20}$	| $f \le
10^{-16}$ \cr
photodissociation of $d$ by photons	| $\beta(m_{10}) \le 10^{-16}$
	| $f \le
10^{-14}$ \cr
distortion of CMBR			| $\beta(m_{13}) \le 10^{-15}$
	| $f \le 10^{-13}$ \cr
photon-to-baryon ratio			| $\beta(m_{13}) \le 10^{-14}$
	| $f \le 10^{-12}$ \cr
$n/p$ by nucleons			| $\beta(m_{10}) \le 10^{-11}$
	| $f \le 10^{-9}$ \cr
entropy production			| $\beta(m_{11}) \le 10^{-3}$
	| $f \le 10^{-1}$ \cr
remnants overclose universe		| $\beta(m_{-2}) \le 10^{-18}$
	| $f \le 10^{-15}$
\endtable
\vskip 0.2in

\noindent{In this table, $m_X$ indicates black holes formed with mass
$10^X g$.
The first four constraints are taken from [\MACGIBBONB].
The limit due to the observed diffuse $\gamma$ ray background at
$100 MeV$ is the strongest constraint.
Uncertainties in
the clustering of black holes and the diffusion of charged particles
within the galaxy may weaken the constraints
on the interstellar $e^\pm$ and $\bar p$ backgrounds.
The next five constraints
are taken from [\LIDSEY]. These constraints focus primarily on the
nucleosynthesis restrictions on black hole radiation.
None of these nucleosynthesis limits are very strong.}

The constraint on black hole remnants requires
that the remnants not overclose the universe.
However, it has also been suggested [\BARROW,\MACGIBBONA] that inert,
Planck mass black hole remnants may provide
a substantal fraction of the dark matter.
Integrating equation III.4, we find
$$\eqalign{\Omega_{bhr}(t_0) = 10^{15} f \le & 1 \cr
\rightarrow f \le & 10^{-15}. \cr }\eqno(IV.2)$$
Black holes of mass $\sim 10^{-2} g$, the first black
holes formed from collapsed cosmic string loops at the
time $t_i$, serve as the dominant source of remnants.
This limit is a conservative upper bound on $f$,
as the energy density in remnants depends critically
on this inital time $t_i$. Although we have
stated earlier that black holes may form only at times
$t \gg t_i$, when the cosmic string loop Schwarschild
radius is much greater than the string thickness,
we have actually included black holes formed starting at time $t_i$.
Therefore, this limit on $f$ may weaken somewhat depending on the
detailed behavior of the collapse of a thick cosmic
string to form a black hole.
Then, closure density in remnants requires $f \approx 10^{-15}$,
which is in disagreement with the  $\gamma$ ray background limit,
$f \le 10^{-17}$.
Therefore, the restrictions on a population of black holes formed
from collapsed cosmic string loops indicate that these
black hole remnants cannot serve as the dark matter.

We may now interpret the restrictions on the population of black
holes
formed from collapsed cosmic string loops in terms of restrictions
on cosmic strings.  The  $\gamma$ ray background limit requires
$$f \le 10^{-17} \quad {\rm for} \quad \mu = 10^{-6}. \eqno(IV.3)$$
This equation gives
the strongest restriction on cosmic strings due to black holes
formed from collapsed cosmic string loops.
If the rough estimates of the magnitude
of $f \sim 10^{-12}$ are reliable, then our results would
indicate that $\mu \le 10^{-11}$
is necessary for compatibility with the  $\gamma$ ray background.
In this case, we could rule out the cosmic string scenario of
large scale structure formation, which demands $\mu \sim 10^{-6}$.
We are not confident in these crude estimates of $f$, however, as
we have indicated in section II.
Clearly, a more detailed investigation is necessary to
definitively determine $f$ [\CG].

\vskip 0.2in
\centerline{\bf V. Conclusion}

In this work we have analyzed the restrictions on
black holes formed from collapsed cosmic string loops.
We have found
that the requirement that the photon flux due to evaporating black
holes
does not exceed the observed  $\gamma$ ray background flux serves as
the
strongest restriction on such black holes formed from cosmic string
loops.
Using a realistic model of cosmic strings,
we find that this observation requires
$f \le 10^{-17}$  for $\mu = 10^{-6}.$
Thus, a fraction of no more than $10^{-17}$ of newly formed cosmic
string loops may collapse to form black holes in order that
$\mu = 10^{-6}$ remains compatible with observation.
This restriction also precludes black hole remnants from serving
as the dark matter.
We plan to study in greater detail the fraction
$f$ of loops which collapse to form black holes [\CG].
If a lower bound $f \ge 10^{-16}$ is found, the  $\gamma$ ray
background limit on evaporating black holes would serve as
the strongest observational bound on cosmic strings,
and would rule out the cosmic string scenario of large-scale
structure formation.

\vskip 0.2in
\centerline{\bf Acknolwedgements}

We would like to thank James E. Lidsey
and Jean Quashnock for useful conversations.
The work of RRC and EG was supported in part by the
DOE (at Chicago and Fermilab) and the
NASA through grant \# NAGW-2381 (at Fermilab).
\par\penalty-400\vskip\chapterskip\spacecheck
   \referenceminspace
   \ifreferenceopen \Closeout\referencewrite \referenceopenfalse \fi
   \line{\fourteenrm\hfil REFERENCES\hfil}\vskip\headskip
   \input referenc.txa

\bye